\documentclass[11pt]{article}
\usepackage{moriond,epsfig}

\def\Journal#1#2#3#4{{#1} {\bf #2}, #3 (#4)}

\def\NIMA{{\em NIM} A}
\def\NPB{{\em Nucl. Phys.} B}
\def\PLB{{\em Phys. Lett.}  B}

\def\PRD{{\em Phys. Rev.} D}

\def\EPJC{{\em Eur. Phys. J.} C}

\def\be{\begin{equation}}
\def\ee{\end{equation}}
\def\bea{\begin{eqnarray}}
\def\eea{\end{eqnarray}}

\begin{document}

\begin{picture}(160,1)
\put(310, 15){\parbox[t]{45mm}{{\tt ROM2F/2002/16}}}
\put(310,  2){\parbox[t]{45mm}{{\tt hep-ex/0206059}}}
\put(-20,-697){---------------------------------------------------} 
\put(-22,-708){ 
{\it {\small Talk presented at the XXXVII$^{\mathrm{th}}$ 
Rencontres de Moriond {\rm``QCD and High Energy Hadronic Interactions''},}} }
\put(-22,-720){ 
{\it {\small Les Arcs, France, 16-23 March 2002.}}
}
\end{picture}

\vspace*{4cm}
\title{SELECTED TOPICS FROM NON-HIGGS SEARCHES AT LEP}

\author{ G. GANIS }

\address{I.N.F.N., Sezione di Roma II,\\
 Dipartimento di Fisica dell' Universit\`a di Roma ``Tor Vergata'',\\ 
 Via Della Ricerca Scientifica 1,\\
 00133 Roma, Italy}

\maketitle\abstracts{
Extensive searches for new phenomena have been performed at LEP. 
The principal aspects and results of those not related to Higgs 
bosons are reviewed here.}

\section{Introduction}
\label{sec:intro}

The last component of the Standard Model (SM) 
yet to be discovered is the Higgs boson. 
Therefore, talking about {\it non-higgs} searches 
implies straightly extensions of the SM. 
This means anything not in the basic SM: new particles, couplings, 
(sub)structure, extra-dimensions, 
new symmetries ... possibly embedded in a coherent and 
complete frame, attempting to solve at least some of the theoretical 
problems of the SM; examples of these frames are those based 
on supersymmetry or technicolor. 
At LEP, both ways of searching for new phenomena could be pursued:
{\it indirect}, via precision measurements, understanding if what is 
measured is consistently described in the SM; {\it direct}, 
disentangling the unexpected by isolating those phenomena most 
likely coming from non-SM terms. 
During 11 years of operation the LEP machine~\cite{lep} delivered about 900 pb$^{-1}$ per 
experiment  at centre-of-mass energies ranging from 88 to 95 GeV (phase LEP~1) and 
130-209 GeV (phase LEP~2). Since the 
four detectors had been designed~\cite{adlo} to do precision physics at the Z peak, 
they featured large covering angle, good particle identification for 
leptons, photons and b quarks, and good jet and energy flow reconstruction.

\vspace{-0.5cm}

\subsection{Precision measurements}
\label{sec:prec}
Precision measurements were performed both at LEP~1 (Z resonance and 
fermion couplings) and at LEP~2 (W boson mass and width, TGC's, anomalous 
couplings). The net result of these measurements is that the SM works 
very well~\cite{lepewwg}; 
constraints are therefore derived to new contributions to the observables. 
As an example, the agreement of the measured Z widths with the SM prediction 
rules out new particles coupling to the Z if their masses are smaller than about 
M$_Z$/2; the sneutrino has been definitely ruled out as 
cold dark matter candidate by this analysis. 
Moreover, the optimal analysis of quantum loop effects suggests that 
Higgs structures with triplets or higher representations are disfavoured 
and that a new SM-like family is excluded at 95\% C.L.~\cite{pdg}; minimal technicolor 
is also disfavoured by precision EW measurements~\cite{pdg}. 

\section{Direct searches}
\label{sec:direct}

Direct searches for new phenomena profit from the highest centre-of-mass 
energy and highest integrated luminosity; therefore they were one of the 
main topics of the highest energy phase of LEP. 
The typical LEP-combined sensitivity for ``5$\sigma$-discovery'' of a 
heavy particle P pair-produced in $\mathrm{e^+e^-}$ collisions is approximated, 
assuming Poisson statistics, 
by $ \sim 10 /(\epsilon_{\mathrm{eff}}\!\cdot\!\int\cal{L}{\mathrm{dt}})$ 
(10 being the minimum number of candidate events to be observed) ;  
with a realistic value for the average effective efficiency 
$\epsilon_{\mathrm{eff}}\sim 0.20$   
this gives, for LEP, $\sim$18 fb if $\mathrm{M_P}\!<$~95 GeV/c$^2$, and   
$\sim$30 fb if 95~$<\!\mathrm{M_P}\!<\sim$~100 GeV/c$^2$.  
Given that the typical SM cross sections are larger than 1~pb, 
{\it the typical LEP~2 potential sensitivity is to couplings an order of magnitude smaller 
of the SM ones or less}. 
This is quantitatively shown, for example, by the results of exotic searches 
like those for compositness and FCNC~\cite{lepexotica}, which parametrize in 
a less model-dependent way possible new effects beyond the SM. 

Moving to more model dependent searches, in the late LEP years growing interest 
was shown for an alternative idea of solving the observed scale hierarchy between 
gravitational and electroweak interactions; the introduction of a certain number n 
of extra-dimensions and of a new fundamental gravitational scale $\mathrm{M_S}$, 
brings new particles in 4-dimensions, called gravitons, and a striking  
{\it single photon} signature in $\mathrm{e^+e^-}$ collisions. LEP sensitivity to 
$\mathrm{M_S}$ decreases with increasing n from about 1.5 TeV at n=2 to about 
0.6 TeV for n=6; the ADLO combination is in progress~\cite{lepexotica}.

\subsection{Supersymmetry}
\label{subsec:susy}

The most popular extensions of the SM are those based on supersymmetry~(SUSY); besides 
providing valuable theoretical benefits~\cite{susy}
 (scale stabilization, a frame for including gravity, 
possible ways of explaining dark matter,...) they are not yet excluded in their 
minimal - and therefore simplest - implementation; moreover, they are in good agreement 
with precision EW data and improve the high energy extrapolation of the theory 
providing a better gauge coupling unification. 
The phenomenological consequences of a Minimal Supersymmetric Model (MSSM) are 
exhaustively described in the literature~\cite{susy}, where definition and notation for 
the relevant parameters can be found.  
LEP had a relevant role in constraining 
the minimal model: some authors claim that about 95\% of the {\it natural} 
parameter space has been strongly disfavoured by LEP results~\cite{strumia}. 
The golden process for discovering SUSY at LEP was chargino  pair production: 
large cross section, clear {\it missing-energy} signature. Difficulties could 
have arisen in peculiar situations: small energy available for chargino decay 
products and abundance of leptonic decays in the case of light sleptons. 
The solution of these problems are a good example of the potentiality of  
LEP detectors: 
the former has been covered exploiting the trigger capabilities of the detectors,
in particular to trigger on low energy isolated photons; 
the latter by taking 
advantage of the good lepton identification, and exploiting for the first 
time the concept of {\it interplay} among searches~\cite{lepmsugra}. 

\begin{figure}
\begin{tabular}{cc}
\epsfig{figure=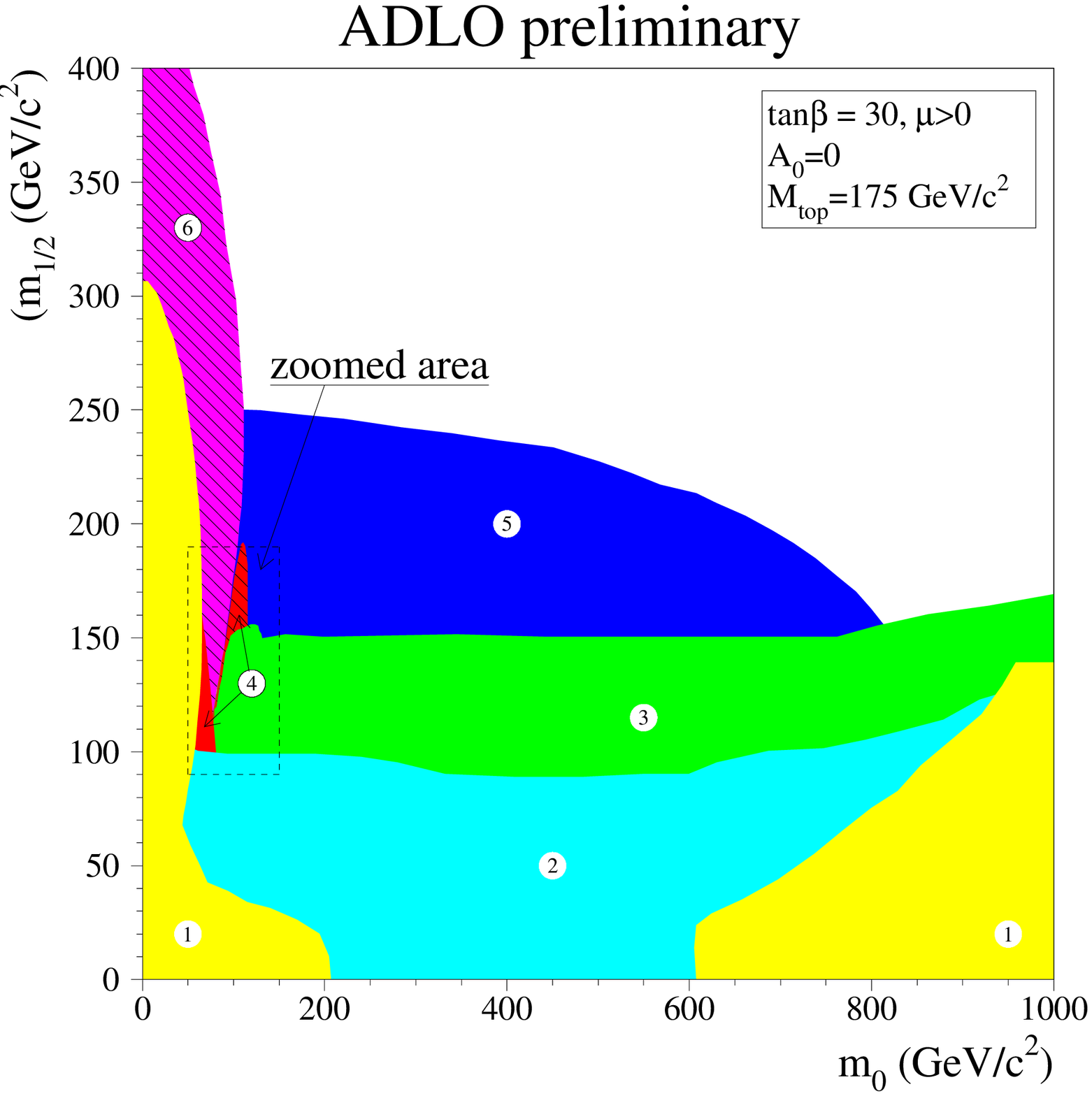,height=7cm} &
\epsfig{figure=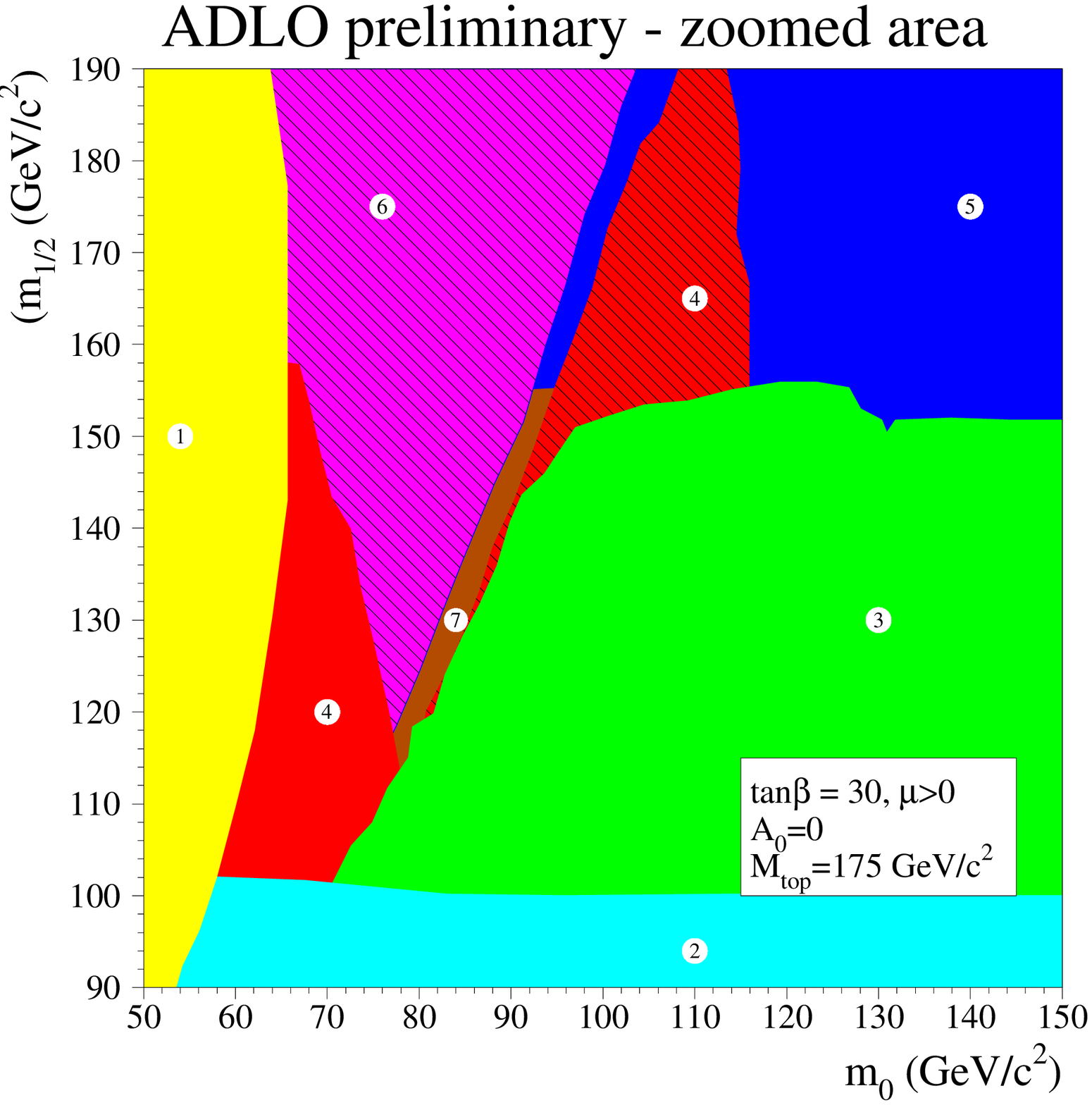,height=7cm} 
\end{tabular}
\caption{Example of interplay in Minimal Supergravity (see text); the 
different regions correspond to: theoretically non-allowed parameter settings (1); 
incompatibility with the Z width measurement at LEP~1 (2); incompatibility with 
LEP~2 searches for charginos (3), selectrons and staus (4), Higgs bosons (5) and
stable staus (6). The zoom shown on the right panel illustrates the impact of special 
analysis looking for 
stau production in heavy neutralino decays when the difference in mass 
between the $\tilde{\tau}$ and the $\chi^0$ is small (7).
}
\label{fig:msugra}
\end{figure}

\begin{figure}
\begin{tabular}{cc}
\epsfig{figure=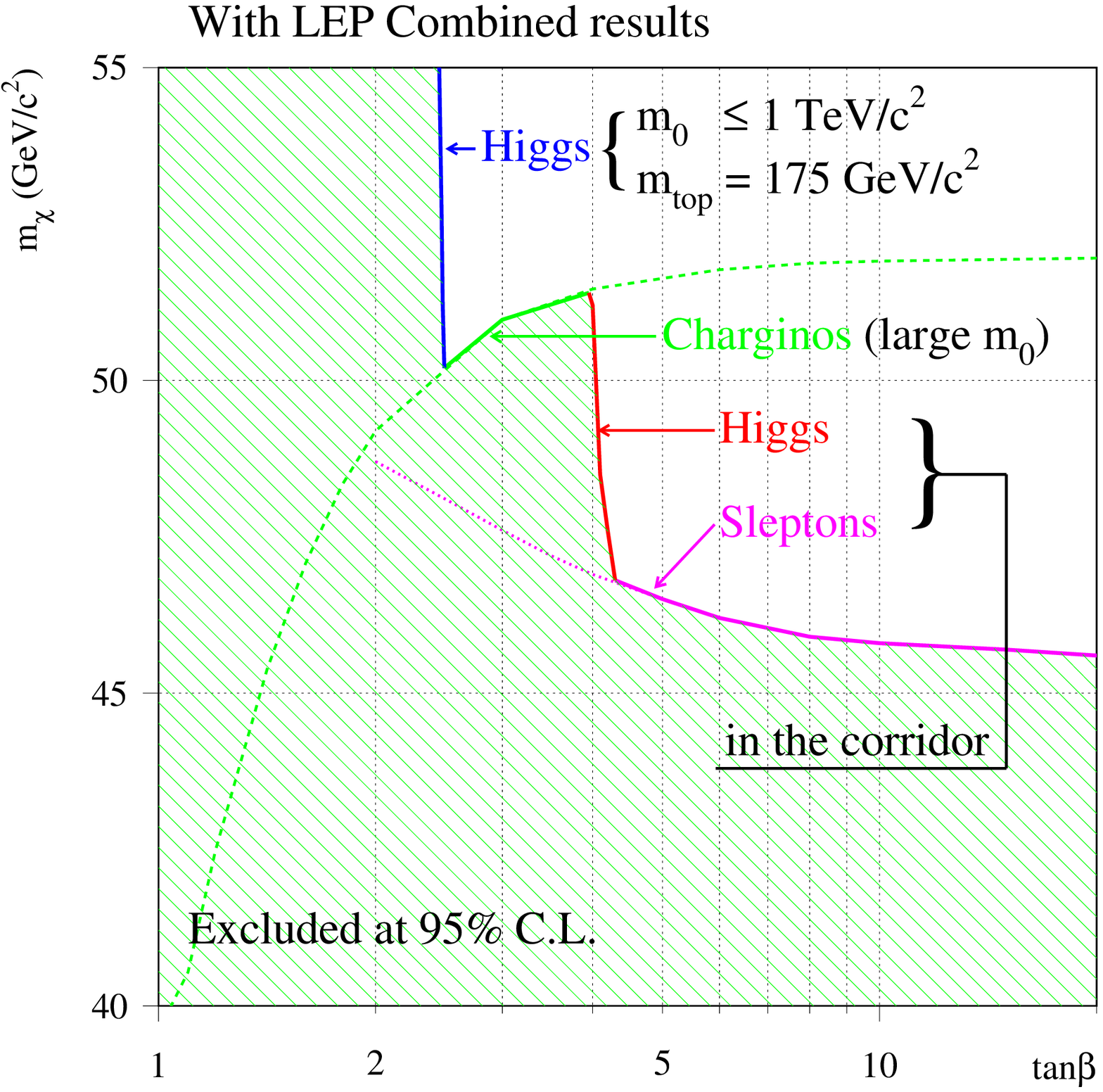,height=7cm} &
\epsfig{figure=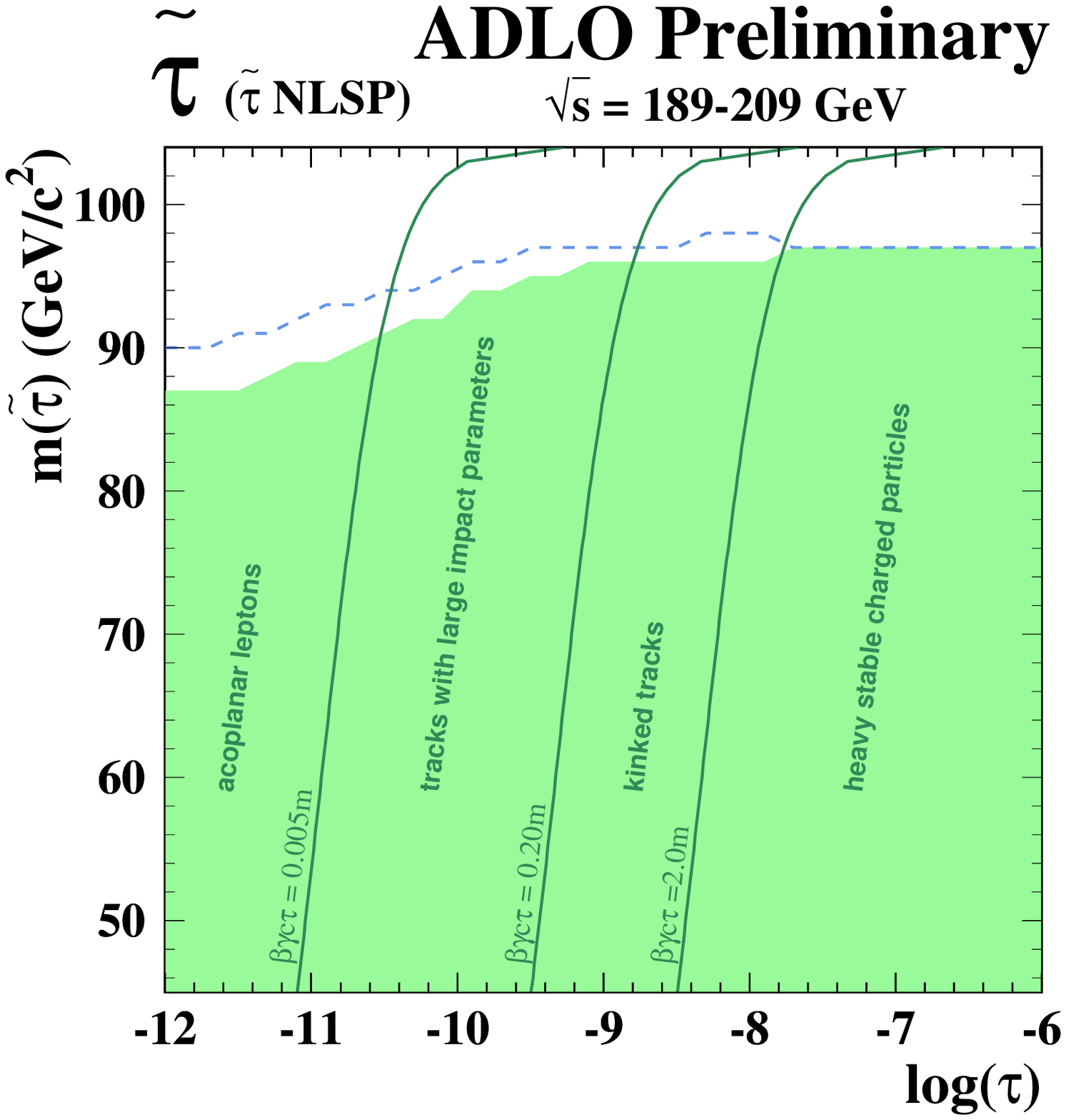,height=7cm} 
\end{tabular}
\caption{Left: Absolute lower limit on the LSP mass in the constrained MSSM
as a function of $\tan\beta$ (see text). 
Right: Lower limit on the stau mass as a function of the lifetime in GMSB 
models (see text). }
\label{fig:mchigmsb}
\end{figure}

The way the different searches interplay among each other is illustrated in 
Fig.~\ref{fig:msugra} 
for a highly constrained MSSM, commonly referred to as Minimal Supergravity~\cite{lepmsugra}; 
the parameter choice here is: $\tan\beta=30$, positive $\mu$, $\mathrm{A_0}$=0 
and the exclusions are shown in the plane of the common sfermion mass $\mathrm{m_0}$ 
and of the common gaugino mass $\mathrm{m_{1/2}}$. 
Notice the significant impact of the Higgs searches even at large values 
of $\tan\beta$. 
Due to stau mixing, pathological situations show-up when the stau is 
almost degenerate with the lightest neutralino: acoplanar tau searches are 
in those cases blind, and the only way to recover sensitivity is to via 
heavy neutralino production and subsequent decay into stau-tau pairs, with at 
least a well energetic tau. These special cases have been addressed by the 
LEP collaborations~\cite{lepmsugra}; the impact is shown in Fig.~\ref{fig:msugra}, right. 
The most famous result of the interplay between searches is the lower limit 
on the lightest neutralino mass, assumed to be the {\it Lighest Supersymmetric 
Particle} (LSP) and a candidate for dark matter if R-parity is 
conserved~\cite{susy}; 
this is shown in Fig.~\ref{fig:mchigmsb}, 
left, in its latest, 
LEP-combined~\cite{lepcmssm}, 
form valid in the so-called {\it constrained} MSSM, where some 
assumptions about unification at very high energy scales are understood~\cite{alephmssm}. 
The LSP must be heavier of about 45 GeV/c$^2$. 
A less famous consequence of LEP results implies that a purely higgsino LSP 
cannot explain the observed amount of cold dark matter~\cite{efgo}. 
The plot in Fig.~\ref{fig:mchigmsb}, left, is obtained after scanning many 
configurations, and subtle but important contributions are hidden under the final 
result. One of these is the absolute lower limit on the stop mass resulting 
from the ALEPH analysis, finalized before this conference~\cite{alephstop} 
and integrated with the analysis of a new stop decay channel, the 4-body  
$\mathrm{\tilde{t}\!\to\!\chi^0bf_u\bar{f}_d}$, of which recently has been 
realized the potential relevance~\cite{celine}.

At the same degree of minimality of the MSSM, there are the so called 
Gauge Mediated Supersymmetric Models (GMSB)~\cite{giudicerattazzi}, 
which differ from the MSSM by the
way supersymmetry breaking (a fundamental ingredient of any realistic 
SUSY model~\cite{susy}) is implemented; the phenomenological novelty
 introduced by GMSB models 
is the possibility of non-negligible lifetime for the Next-to-LSP, which 
typically is either the neutralino or the stau. These particle could therefore
decay with the tracking volumes of the LEP detectors leading to kink or 
detached-vertex topologies. The good sensitivity of all LEP detectors to this 
kind of topologies has allowed to produce lifetime independent constrains 
on the NLSP particle. Figure~\ref{fig:mchigmsb}, right, shows the result 
in the case of the stau, as obtained from an ADLO combination~\cite{lepgmsb}
presented for the first time at this conference.

Finally, as an example of non-minimal SUSY searches, it is worth mentioning 
the analyses performed allowing R-parity violation terms in the theory. In this 
case the LSP can decay leading to a plethora of possible final states, with 
many leptons and/or quarks, with or without missing-energy. The many analyses 
(whose results are now being LEP-combined~\cite{leprpv}) have shown that 
the discovery of supersymmetry was not missed because of R-parity violation. 
Most significantly, the study of these decays allowed to test very peculiar 
topologies, otherwise unsearched for. 

\section{Conclusions}
\label{sec:conclu}

The LEP impact on possible extensions of the SM has been quite 
significant. Precision measurements at the EW scale proved the 
basic correctness of the SM, disfavouring, among others, heavy Higgs bosons, 
complicated Higgs structures, additional copies of the known 
families, QCD-like technicolor frames. 
Unsuccessful direct searches proved sensitivities to couplings
a order of magnitude smaller than the  
SM ones and typically excluded new particle with masses 
smaller than about 100 GeV/c$^2$. 
The interplay among several searches could be used to constrain 
well-defined minimal extensions of the SM, so allowing to derive 
absolute limits on some parameters, like the lower limit on the 
LSP mass in the constrained MSSM. 
For completeness and solidity of its results, LEP has certainly set 
a new standard  for future searches.

\vspace{-0.3cm}

\paragraph{Acknowledgments}
I would like to thank the contact persons of the relevant LEP working 
groups for promptly providing material and/or support in the preparation 
of this talk, and G. Sguazzoni for the careful reading of the manuscript.

\vspace{-0.3cm}

\paragraph{References}

%\vspace{-0.5cm}


\begin{thebibliography}{99}
\bibitem{lep} See, for example, R.~Assmann, Proceedings of Chamonix 2001, CERN-SL-2001-003~DI 

\bibitem{adlo} ALEPH: \Journal{\NIMA}{294}{121}{1990}; 
               DELPHI: \Journal{\NIMA}{303}{233}{1991}; 
   L3: \Journal{\NIMA}{289}{35}{1990}; OPAL: \Journal{\NIMA}{305}{275}{1991}.

\bibitem{lepewwg} {\tt http://lepewwg.web.cern.ch/LEPEWWG/}

\bibitem{pdg} Particle Data Group, \Journal{\EPJC}{15}{1}{2000}

\bibitem{lepexotica} {\tt http://lepexotica.web.cern.ch/LEPEXOTICA/}

\bibitem{susy} See, for example, S.~Martin, ``A Supersymmetry primer'', 
%in {\it Perspectives on supersymmetry}, 1-98, Kane, G.L. (ed.); 
{\tt hep-ph/9709356}; also, N. Polonsky, ``Supersymmetry: Structure and Phenomena'',
 Lect. Notes Phys. M68, 1-169 (2001),  {\tt hep-ph/0108236}.  

%\bibitem{susy} See, for example, S.~Martin, ``A Supersymmetry primer'', 
%in {\it Perspectives on supersymmetry}, 1-98, Kane, G.L. (ed.); 
%{\tt hep-ph/9709356}. 

\bibitem{strumia} See, for example, L.~Giusti, A.~Romanino and A.~Strumia, \Journal{\NPB} 
{550}{3}{1999}. 

\bibitem{lepmsugra} 
LEPSUSYWG, ALEPH, DELPHI, L3 and OPAL exps, note LEPSUSYWG/02-06.1 . 

\bibitem{lepcmssm} 
LEPSUSYWG, ALEPH, DELPHI, L3 and OPAL exps, note LEPSUSYWG/01-07.1 . 

\bibitem{alephmssm} See ALEPH Coll., \Journal{\PLB}{499}{67}{2001} for a detailed explanation 
     of Fig.~2, left. 

\bibitem{efgo} J.~Ellis at al., \Journal{\PRD}{62}{075010}{2000}

\bibitem{alephstop} ALEPH Coll., \Journal{\PLB}{537}{5}{2002} 

\bibitem{celine} C.~B\"ohm,A.~Djouadi and Y.~Mambrini, \Journal{\PRD}{61}{095006}{2000}.

\bibitem{giudicerattazzi} G.~Giudice and R.~Rattazzi, {\em Phys. Rept.} {\bf 322}, 419 (1999).

\bibitem{lepgmsb} 
LEPSUSYWG, ALEPH, DELPHI, L3 and OPAL exps, note LEPSUSYWG/02-09.1 . 

\bibitem{leprpv} 
LEPSUSYWG, ALEPH, DELPHI, L3 and OPAL exps, note LEPSUSYWG/02-10.1 . 

\end{thebibliography}
\end{document}